\documentclass[conference]{IEEEtran}
\usepackage[T1]{fontenc}
\usepackage{newtxtext,newtxmath}
\usepackage{graphicx}
\usepackage{amsmath,amssymb}
\usepackage{bm}
\usepackage{cite}
\usepackage{booktabs}
\usepackage{array}
\usepackage{hyperref}
\usepackage{siunitx}
\usepackage{mathtools}
\hypersetup{hidelinks}
\begin{document}
\title{FEM-Based Dispersion and Mode Analysis of Rectangular, Circular, and Ridge Waveguide Geometries}

\author{
\IEEEauthorblockN{Sabrina Saima}
\IEEEauthorblockA{
Elmore Family School of Electrical and Computer Engineering\\
Purdue University\\
West Lafayette, IN, USA\\
ssaima@purdue.edu}
}

\maketitle

\begin{abstract}
This paper presents a two-dimensional finite element method (FEM) solver for computing modal field distributions and dispersion characteristics of hollow metallic waveguides. To solve the waveguide problem, the source-free frequency-domain Maxwell equations are reduced to scalar Helmholtz eigenvalue formulations evaluated over the waveguide's transverse cross section. The computational method determines both transverse electric (TE) and transverse magnetic (TM) mode families by enforcing perfectly electrically conducting (PEC) boundary conditions. The framework is initially validated against analytical benchmarks using empty rectangular and circular waveguides, demonstrating high accuracy in computing cutoff wavenumbers, dispersion curves, and field maps for the first three unique modes. After validation, the solver is applied to analyze single-ridged and double-ridged waveguides. The numerical results demonstrate that introducing metallic ridges successfully redistributes the modal fields and significantly lowers the cutoff frequency of the dominant mode relative to empty rectangular guides. Ultimately, this work confirms that the generalized eigenvalue FEM formulation is a robust and adaptable tool for analyzing complex waveguide geometries where exact analytical solutions are unavailable.
\end{abstract}


\section{Introduction}

Waveguides are fundamental structures in microwave and electromagnetic engineering because they confine and guide electromagnetic energy with relatively low loss. Among the classical examples, hollow rectangular and circular waveguides are particularly important because their modal fields and cutoff frequencies can be obtained analytically, making them standard benchmark problems in computational electromagnetics \cite{pozar,collin}. These canonical structures are widely used to validate numerical solvers before applying them to more complicated geometries.

The behavior of a waveguide is governed by its modal content. Each mode has a specific field distribution and a cutoff frequency below which propagation cannot occur. For ideal rectangular and circular waveguides, the modal solutions can be derived from Maxwell's equations using separation of variables, leading to the familiar TE and TM mode families \cite{cheng,pozar,sadiku}. The corresponding dispersion relations describe how the propagation constant varies with frequency and determine the useful operating band of the guide. In ridged waveguides, the introduction of one or two metallic ridges modifies the boundary conditions and redistributes the electric field, lowering the cutoff frequency of the dominant mode and changing the separation between modes \cite{collin}.

A number of numerical methods may be used to study waveguides when analytical solutions are not available. The finite-difference time-domain (FDTD) method is popular for time-domain and broadband analysis because it directly advances the fields on a grid \cite{fdtdfemmom,jin_field}; it is also widely used in applied electromagnetic and photonic device modeling, including resonator-based sensing \cite{intisar2024soi}. The method of moments (MoM) is often effective for radiation and scattering problems involving conductors \cite{harrington1993}, and for electrically large problems fast algorithms such as the fast multipole method can be used to accelerate integral-equation computations \cite{greengard1987,jin_field}. However, for modal analysis of hollow waveguides with complicated cross sections, these methods are less convenient than the finite element method (FEM). In particular, FEM is well suited to structures with curved or irregular boundaries because it uses an unstructured mesh and can represent complex geometry accurately \cite{jin_fem}. It is also naturally formulated as an eigenvalue problem on a two-dimensional transverse cross section, which makes it especially appropriate for calculating waveguide cutoff frequencies, modal fields, and dispersion curves \cite{jin_fem}. FEM-based modeling is also common in photonics, including waveguide and subwavelength device studies \cite{saima2023mim,zahin2026pra}. For these reasons, FEM is the method chosen in this work.

A two-dimensional FEM eigenmode solver is developed to compute modal field profiles and dispersion for rectangular, circular, single-ridged, and double-ridged waveguides. Rectangular and circular waveguides are used for validation against analytical cutoff frequencies and mode patterns \cite{pozar,cheng}, after which the same formulation is applied to ridged waveguides to quantify the impact of ridge loading on cutoff and dispersion.

The remainder of the paper is organized as follows: Section~II presents the governing equations and FEM formulation; Section~III summarizes the geometry, mesh, and simulation parameters; Section~IV reports the numerical results; and Section~V concludes with possible extensions.

\section{Formulation and Discretization}

\subsection{Maxwell's Equations}

We begin with the source-free Maxwell curl equations in a linear, isotropic, homogeneous medium in the frequency domain:
\begin{align}
\nabla \times \mathbf{E} &= -j\omega\mu \mathbf{H}, \label{eq:maxwell1}\\
\nabla \times \mathbf{H} &= \phantom{-}j\omega\epsilon \mathbf{E}. \label{eq:maxwell2}
\end{align}

The waveguides considered here are uniform along the $z$-direction, infinitely long, and filled with a homogeneous medium. Hence, the fields can be expressed as
\begin{align}
\mathbf{E}(x,y,z) &= \left[\mathbf{E}_t(x,y) + \hat{\mathbf{z}}\,E_z(x,y)\right]e^{-j\beta z}, \label{eq:E_decomp}\\
\mathbf{H}(x,y,z) &= \left[\mathbf{H}_t(x,y) + \hat{\mathbf{z}}\,H_z(x,y)\right]e^{-j\beta z}, \label{eq:H_decomp}
\end{align}
where $\beta$ is the propagation constant, $\mathbf{E}_t$ and $\mathbf{H}_t$ are the transverse field components, and $E_z$ and $H_z$ are the longitudinal field components.

Using the assumed $e^{-j\beta z}$ dependence, the Helmholtz equation reduces to
\begin{equation}
\nabla_t^2 \mathbf{E} + (k^2-\beta^2)\mathbf{E} = 0,
\label{eq:transverse_wave}
\end{equation}
where
\begin{equation}
k=\omega\sqrt{\mu\epsilon}
\end{equation}
is the wavenumber of the filling medium. It is convenient to define the cutoff wavenumber as
\begin{equation}
k_c^2 = k^2-\beta^2.
\label{eq:kc_def}
\end{equation}
Therefore, the dispersion relation is
\begin{equation}
\beta^2 = k^2-k_c^2.
\label{eq:dispersion_relation}
\end{equation}
Equation \eqref{eq:dispersion_relation} shows that propagation occurs only when $k > k_c$. Otherwise, $\beta$ becomes purely imaginary and the mode is below cutoff.

\subsection{TE and TM Scalar Formulations}

For a homogeneous waveguide, the longitudinal field components decouple and the modal solutions separate into the usual TE and TM families. In particular, TM modes satisfy $H_z=0$ with $E_z\neq 0$, whereas TE modes satisfy $E_z=0$ with $H_z\neq 0$. Thus, each modal family can be obtained from a scalar eigenvalue problem posed over the two-dimensional waveguide cross section.

\subsubsection{TM modes}

For TM modes, the unknown longitudinal scalar field is $E_z(x,y)$, and it satisfies
\begin{equation}
\nabla_t^2 E_z + k_c^2 E_z = 0 \qquad \text{in } \Omega.
\label{eq:tm_scalar_pde}
\end{equation}
For a perfectly electrically conducting (PEC) boundary, the tangential electric field must vanish. Since $E_z$ is tangential to the PEC wall, the boundary condition is
\begin{equation}
E_z = 0 \qquad \text{on } \Gamma.
\label{eq:tm_bc}
\end{equation}

\subsubsection{TE modes}

For TE modes, the unknown longitudinal scalar field is $H_z(x,y)$, and it satisfies
\begin{equation}
\nabla_t^2 H_z + k_c^2 H_z = 0 \qquad \text{in } \Omega.
\label{eq:te_scalar_pde}
\end{equation}
For a PEC boundary, the corresponding boundary condition becomes
\begin{equation}
\frac{\partial H_z}{\partial n}=0 \qquad \text{on } \Gamma,
\label{eq:te_bc}
\end{equation}
where $\partial/\partial n$ denotes differentiation along the outward normal direction.

\subsection{Why This is an Eigenvalue Problem}

At this stage, the important point is that the field distribution and the cutoff wavenumber are both unknown. For either TE or TM modes, let the scalar modal field be denoted by
\[
u=
\begin{cases}
E_z, & \text{TM modes},\\
H_z, & \text{TE modes}.
\end{cases}
\]
Then both cases can be written compactly as
\begin{equation}
\nabla_t^2 u + k_c^2 u = 0 \qquad \text{in } \Omega.
\label{eq:generic_scalar}
\end{equation}

This has the same mathematical structure as a standard eigenvalue problem:
\begin{equation}
\mathcal{L}u = \lambda u,
\end{equation}
where the differential operator $\mathcal{L}$ plays the role of the matrix operator, the modal field $u$ is the eigenfunction, and the scalar $k_c^2$ is the eigenvalue. Therefore, in this waveguide problem:
\begin{itemize}
    \item the \textbf{eigenvector/eigenfunction} is the modal field distribution $u_m$,
    \item the \textbf{eigenvalue} is the cutoff wavenumber squared, $k_{c,m}^2$.
\end{itemize}

Once $k_{c,m}$ is known, the cutoff frequency of the $m$th mode is
\begin{equation}
f_{c,m}=\frac{k_{c,m}}{2\pi\sqrt{\mu\epsilon}}.
\label{eq:fc_general}
\end{equation}
For the empty waveguides considered in this work, this reduces to
\begin{equation}
f_{c,m}=\frac{c_0\,k_{c,m}}{2\pi},
\label{eq:fc_vacuum}
\end{equation}
where $c_0$ is the speed of light in free space.

\subsection{Weak Form}

To derive the finite element formulation, \eqref{eq:generic_scalar} is tested with a weighting function $w$ and integrated over the cross section $\Omega$:
\begin{equation}
\int_{\Omega} w\left(\nabla_t^2 u + k_c^2 u\right)d\Omega = 0.
\label{eq:tested_form}
\end{equation}
Applying integration by parts gives
\begin{equation}
-\int_{\Omega}\nabla_t w\cdot\nabla_t u\,d\Omega
+ \int_{\Gamma} w\,\frac{\partial u}{\partial n}\,d\Gamma
+ k_c^2\int_{\Omega}wu\,d\Omega = 0.
\label{eq:ibp_form}
\end{equation}
Rearranging,
\begin{equation}
\int_{\Omega}\nabla_t w\cdot\nabla_t u\,d\Omega
=
k_c^2\int_{\Omega}wu\,d\Omega
+
\int_{\Gamma} w\,\frac{\partial u}{\partial n}\,d\Gamma.
\label{eq:weak_general}
\end{equation}

The boundary term is then handled according to the modal family:
\begin{itemize}
    \item For \textbf{TM modes}, the Dirichlet condition $E_z=0$ is enforced strongly by removing boundary nodes from the unknown set. Hence the testing and trial functions are taken only over interior nodes, and the boundary contribution does not appear in the final system.
    \item For \textbf{TE modes}, the Neumann condition $\partial H_z/\partial n=0$ makes the boundary term vanish naturally.
\end{itemize}

Therefore, in both cases the final weak form becomes
\begin{equation}
\int_{\Omega}\nabla_t w\cdot\nabla_t u\,d\Omega
=
k_c^2\int_{\Omega}wu\,d\Omega.
\label{eq:weak_final}
\end{equation}

\subsection{Finite Element Approximation}

The waveguide cross section is discretized using first-order triangular elements. Let $N_i(x,y)$ denote the standard linear nodal basis functions. Then the scalar field is approximated as
\begin{equation}
u(x,y) \approx \sum_{j=1}^{N_u} U_j N_j(x,y),
\label{eq:fem_expansion}
\end{equation}
where $U_j$ are the unknown nodal coefficients and $N_u$ is the number of unknown nodes retained after applying the boundary-condition treatment.

Using the Galerkin method, the testing functions are chosen from the same basis:
\begin{equation}
w=N_i.
\end{equation}
Substituting \eqref{eq:fem_expansion} into \eqref{eq:weak_final} gives
\begin{equation}
\sum_{j=1}^{N_u} U_j
\int_{\Omega}\nabla_t N_i\cdot\nabla_t N_j\,d\Omega
=
k_c^2
\sum_{j=1}^{N_u} U_j
\int_{\Omega}N_iN_j\,d\Omega.
\label{eq:before_matrix}
\end{equation}

\subsection{Definition of the Global Matrices}

Equation \eqref{eq:before_matrix} leads directly to the generalized matrix eigenvalue problem
\begin{equation}
[A]\{U_m\} = k_{c,m}^2 [B]\{U_m\}.
\label{eq:global_eig}
\end{equation}
Here,
\begin{equation}
[A]_{ij} = \int_{\Omega}\nabla_t N_i\cdot\nabla_t N_j\,d\Omega,
\label{eq:A_def}
\end{equation}
and
\begin{equation}
[B]_{ij} = \int_{\Omega}N_iN_j\,d\Omega.
\label{eq:B_def}
\end{equation}

The matrix $[A]$ is the \textbf{stiffness matrix}, since it comes from the gradient-gradient term of the weak form. The matrix $[B]$ is the \textbf{mass matrix}, since it comes from the inner-product term involving the field itself. 
Thus, the system being solved numerically is not a forced linear system like $[K]\{x\}=\{b\}$, but rather the generalized eigenvalue problem \eqref{eq:global_eig}. The eigenvector $\{U_m\}$ gives the sampled modal field shape, and the eigenvalue $k_{c,m}^2$ gives the cutoff wavenumber squared of that mode.

\subsection{Element-Level Discretization and Assembly}

Within each triangular element $\Omega_e$, the field is approximated by
\begin{equation}
u^{(e)}(x,y)=\sum_{j=1}^{3}u_j^{(e)}N_j^{(e)}(x,y),
\label{eq:local_interp}
\end{equation}
where $N_j^{(e)}$ are the local linear basis functions. Since linear basis functions are used, their gradients are constant over each element. Therefore, the element matrices are
\begin{align}
[A^{(e)}]_{ij} &= \int_{\Omega_e}\nabla_t N_i^{(e)}\cdot\nabla_t N_j^{(e)}\,d\Omega,
\label{eq:elem_stiff}\\
[B^{(e)}]_{ij} &= \int_{\Omega_e}N_i^{(e)}N_j^{(e)}\,d\Omega.
\label{eq:elem_mass}
\end{align}
These element-level contributions are assembled into the global matrices $[A]$ and $[B]$ using the element connectivity list. Because each basis function has compact support, the resulting global matrices are sparse.

\subsection{Treatment of Boundary Conditions in the Final Matrix System}

The boundary treatment determines exactly which final matrix equation is solved.

\subsubsection{TM case}

For TM modes, the PEC condition
\[
E_z=0 \qquad \text{on } \Gamma
\]
is enforced strongly. Hence, all boundary nodes are removed from the unknown vector, and the final generalized eigenvalue problem is assembled only over the interior nodes:
\begin{equation}
[A_{\mathrm{int}}]\{U_m^{\mathrm{TM}}\}
=
k_{c,m}^2
[B_{\mathrm{int}}]\{U_m^{\mathrm{TM}}\}.
\label{eq:tm_final_eig}
\end{equation}

\subsubsection{TE case}

For TE modes, the homogeneous Neumann condition
\[
\frac{\partial H_z}{\partial n}=0 \qquad \text{on } \Gamma
\]
is natural, so boundary nodes remain in the system. Therefore, the TE modes are obtained from
\begin{equation}
[A]\{U_m^{\mathrm{TE}}\}
=
k_{c,m}^2
[B]\{U_m^{\mathrm{TE}}\}.
\label{eq:te_final_eig}
\end{equation}
Because of the Neumann boundary condition, a trivial zero eigenvalue corresponding to a constant field may appear. This non-physical solution is discarded, and the first nonzero eigenvalues are taken as the physical TE cutoff modes.

\subsection{Recovered Quantities and Mode Visualization}

After solving the generalized eigenvalue problem, the eigenpairs
\[
\left(k_{c,m}^2,\{U_m\}\right)
\]
are obtained. From these:
\begin{itemize}
    \item $k_{c,m}=\sqrt{k_{c,m}^2}$ gives the cutoff wavenumber,
    \item $f_{c,m}$ is computed using \eqref{eq:fc_general} or \eqref{eq:fc_vacuum},
    \item the modal field distribution is given by the eigenvector $\{U_m\}$.
\end{itemize}

For visualization of transverse electric-field patterns in the cross section, the longitudinal solution is post-processed as
\begin{equation}
\mathbf{E}_t \propto -\nabla_t E_z \qquad \text{for TM modes},
\label{eq:tm_et}
\end{equation}
and
\begin{equation}
\mathbf{E}_t \propto \hat{\mathbf{z}}\times\nabla_t H_z \qquad \text{for TE modes}.
\label{eq:te_et}
\end{equation}
These relations are sufficient for plotting the modal field directions and comparing the numerically obtained modes with the expected analytical mode shapes.

\section{Simulation Setup}

All simulations were performed using MATLAB implementation of the scalar finite element method described in the previous section. The waveguides were modeled as empty, perfectly electrically conducting (PEC) structures that are uniform and infinitely long in the longitudinal direction. Modal analysis was carried out on the two-dimensional transverse cross section of each guide. In all cases, the first three unique TE modes and the first three unique TM modes were retained for dispersion and field-pattern analysis.

The physical constants used throughout the study were the free-space values $c_0 = 3\times 10^8$~m/s and $\mu_0 = 4\pi\times 10^{-7}$~H/m, with $\epsilon_0 = 1/(\mu_0 c_0^2)$. The geometric and meshing parameters for the four waveguide configurations are summarized in Table~\ref{tab:geometry_setup}.

\begin{table}[t]
\centering
\caption{Waveguide geometry and mesh parameters}
\label{tab:geometry_setup}
\begin{tabular}{p{2.5cm} p{5.1cm}}
\toprule
Geometry & Parameters \\
\midrule
Rectangular guide &
Outer dimensions: $a = 0.08$ m, $b = 0.04$ m \\
&
Aspect ratio: $a:b = 2:1$ \\
&
Structured mesh divisions: $n_x = 38$, $n_y = 19$ \\
&
Each rectangular cell split into two triangular elements \\[3pt]

Circular guide &
Radius: $R = 0.03$ m \\
&
Unstructured triangular mesh \\
&
Mesh-size parameter: $h \approx R/15.5$ \\[3pt]

Single-ridged guide &
Outer dimensions: $a = 0.08$ m, $b = 0.04$ m \\
&
Ridge span: $0.025 \le x \le 0.055$ m \\
&
Ridge width: $w_r = 0.030$ m \\
&
Ridge height: $h_r = 0.010$ m \\[3pt]

Double-ridged guide &
Outer dimensions: $a = 0.08$ m, $b = 0.04$ m \\
&
Top and bottom ridge span: $0.025 \le x \le 0.055$ m \\
&
Ridge width: $w_d = 0.030$ m \\
&
Each ridge height: $h_d = 0.010$ m \\
\bottomrule
\end{tabular}
\end{table}

For the rectangular guide, the dispersion curves were plotted in normalized form using $k_0 a$ as the horizontal axis. For the circular guide, the corresponding normalization used $k_0 R$. For the ridged waveguides, the dispersion curves were also reported in normalized form using the outer rectangular width as the reference length. In addition to the dispersion plots, scalar modal fields and reconstructed transverse electric field patterns were generated for each geometry.

\section{Numerical Results}
The rectangular and circular waveguides were used first as validation cases because analytical modal solutions are available for both geometries. These two cases therefore provide a direct check of the FEM-computed cutoff values, dispersion curves, and modal field patterns before extending the same formulation to the single-ridged and double-ridged waveguides.
\subsection{Validation for the Rectangular Waveguide}
The analytical transverse field expressions used for comparison were the standard rectangular-waveguide field components \cite{pozar}. For the TEmn modes, the transverse fields are
\begin{align}
E_x &= \frac{j\omega\mu n\pi}{k_c^2 b}\,
A_{mn}\cos\frac{m\pi x}{a}\sin\frac{n\pi y}{b}\,e^{-j\beta z},
\label{eq:rect_te_ex_num}\\[4pt]
E_y &= -\frac{j\omega\mu m\pi}{k_c^2 a}\,
A_{mn}\sin\frac{m\pi x}{a}\cos\frac{n\pi y}{b}\,e^{-j\beta z},
\label{eq:rect_te_ey_num}\\[4pt]
H_x &= \frac{j\beta m\pi}{k_c^2 a}\,
A_{mn}\sin\frac{m\pi x}{a}\cos\frac{n\pi y}{b}\,e^{-j\beta z},
\label{eq:rect_te_hx_num}\\[4pt]
H_y &= \frac{j\beta n\pi}{k_c^2 b}\,
A_{mn}\cos\frac{m\pi x}{a}\sin\frac{n\pi y}{b}\,e^{-j\beta z},
\label{eq:rect_te_hy_num}
\end{align}
while for the TMmn modes, the transverse fields are
\begin{align}
E_x &= -\frac{j\beta m\pi}{a k_c^2}\,
B_{mn}\cos\frac{m\pi x}{a}\sin\frac{n\pi y}{b}\,e^{-j\beta z},
\label{eq:rect_tm_ex_num}\\[4pt]
E_y &= -\frac{j\beta n\pi}{b k_c^2}\,
B_{mn}\sin\frac{m\pi x}{a}\cos\frac{n\pi y}{b}\,e^{-j\beta z},
\label{eq:rect_tm_ey_num}\\[4pt]
H_x &= \frac{j\omega\epsilon n\pi}{b k_c^2}\,
B_{mn}\sin\frac{m\pi x}{a}\cos\frac{n\pi y}{b}\,e^{-j\beta z},
\label{eq:rect_tm_hx_num}\\[4pt]
H_y &= -\frac{j\omega\epsilon m\pi}{a k_c^2}\,
B_{mn}\cos\frac{m\pi x}{a}\sin\frac{n\pi y}{b}\,e^{-j\beta z}.
\label{eq:rect_tm_hy_num}
\end{align}
In this geometry, the first step was to compare the computed cutoff values against the standard rectangular-waveguide formulas and then verify that the field patterns and dispersion behavior followed the expected modal ordering.

For the hollow rectangular waveguide, the analytical cutoff relation used for comparison was
\begin{equation}
k_{c,mn}
=
\sqrt{
\left(\frac{m\pi}{a}\right)^2
+
\left(\frac{n\pi}{b}\right)^2
},
\label{eq:rect_kc_num}
\end{equation}
with the corresponding cutoff frequency
\begin{equation}
f_{c,mn}
=
\frac{1}{2\pi\sqrt{\mu\epsilon}}
\sqrt{
\left(\frac{m\pi}{a}\right)^2
+
\left(\frac{n\pi}{b}\right)^2
}.
\label{eq:rect_fc_num}
\end{equation}

From \eqref{eq:rect_fc_num}, the dominant TE mode is TE$_{10}$, with cutoff frequency
\begin{equation}
f_{c,10}^{(\mathrm{TE})}
=
\frac{1}{2a\sqrt{\mu\epsilon}}.
\label{eq:rect_te10_fc}
\end{equation}
This is the first cutoff mode because $a>b$, so the quantity $\pi/a$ is smaller than $\pi/b$, making TE$_{10}$ lower in cutoff than TE$_{01}$ or TE$_{11}$. There is no TE$_{00}$ mode because if $m=n=0$, the field expressions vanish identically. For TM modes, neither index can be zero, so TM$_{00}$, TM$_{10}$, and TM$_{01}$ do not exist. As a result, the first TM cutoff mode is TM$_{11}$, with cutoff frequency
\begin{equation}
f_{c,11}^{(\mathrm{TM})}
=
\frac{1}{2\pi\sqrt{\mu\epsilon}}
\sqrt{
\left(\frac{\pi}{a}\right)^2
+
\left(\frac{\pi}{b}\right)^2 }.
\label{eq:rect_tm11_fc}
\end{equation}

In the numerical implementation, the rectangular guide was the simplest case because the outer boundary coincides exactly with the mesh lines. A structured nodal grid was first created over the cross section, and each rectangular cell was then split into two triangular elements for the FEM assembly. An additional post-processing step was used to avoid repeated reporting of nearly identical cutoff values. In this context, a \emph{degenerate} case refers to two modes having the same, or nearly the same, cutoff eigenvalue. Even when the exact analytical values are not identical, the numerical solver can still return very closely spaced roots corresponding to repeated or nearly repeated modes. To avoid plotting redundant branches, the code retained only the first three \emph{unique} TE modes and the first three \emph{unique} TM modes.

Table~\ref{tab:rectangular_cutoff_comparison} compares the analytical and FEM cutoff values for the first three unique TE modes and the first three unique TM modes of the rectangular waveguide.

\begin{table}[t]
\centering
\caption{Rectangular-waveguide cutoff comparison}
\label{tab:rectangular_cutoff_comparison}
\begin{tabular}{ccccc}
\toprule
Mode & Analytical $k_c$ & FEM $k_c$ & Analytical $f_c$ & FEM $f_c$ \\
     & (rad/m) & (rad/m) & (GHz) & (GHz) \\
\midrule
TE$_{10}$ & 39.269908 & 39.281064 & 1.875000 & 1.875533 \\
TE$_{20}$ & 78.539816 & 78.629010 & 3.750000 & 3.754259 \\
TE$_{11}$ & 87.810184 & 87.974695 & 4.192627 & 4.200482 \\
TM$_{11}$ & 87.810184 & 87.975218 & 4.192627 & 4.200507 \\
TM$_{21}$ & 111.072073 & 111.451296 & 5.303301 & 5.321407 \\
TM$_{31}$ & 141.589667 & 142.335258 & 6.760409 & 6.796008 \\
\bottomrule
\end{tabular}
\end{table}

\begin{figure}
    \centering
     \includegraphics[width=0.90\linewidth]{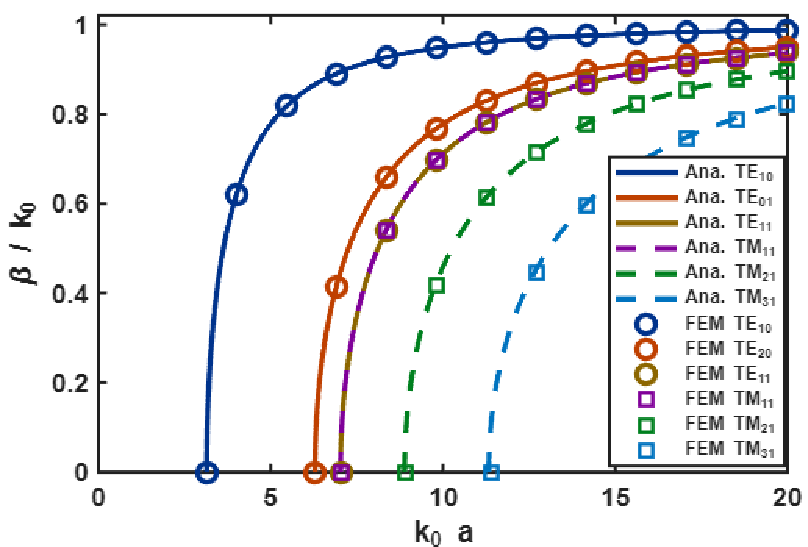}
    \caption{Analytical and FEM dispersion curves for the rectangular waveguide.}
    \label{fig:rect_dispersion}
\end{figure}

Figure~\ref{fig:rect_dispersion} compares the analytical and FEM dispersion curves for the retained rectangular-waveguide modes. The agreement is close, and the FEM branches follow the expected analytical behavior well.

\begin{figure}
    \centering
     \includegraphics[width=\linewidth]{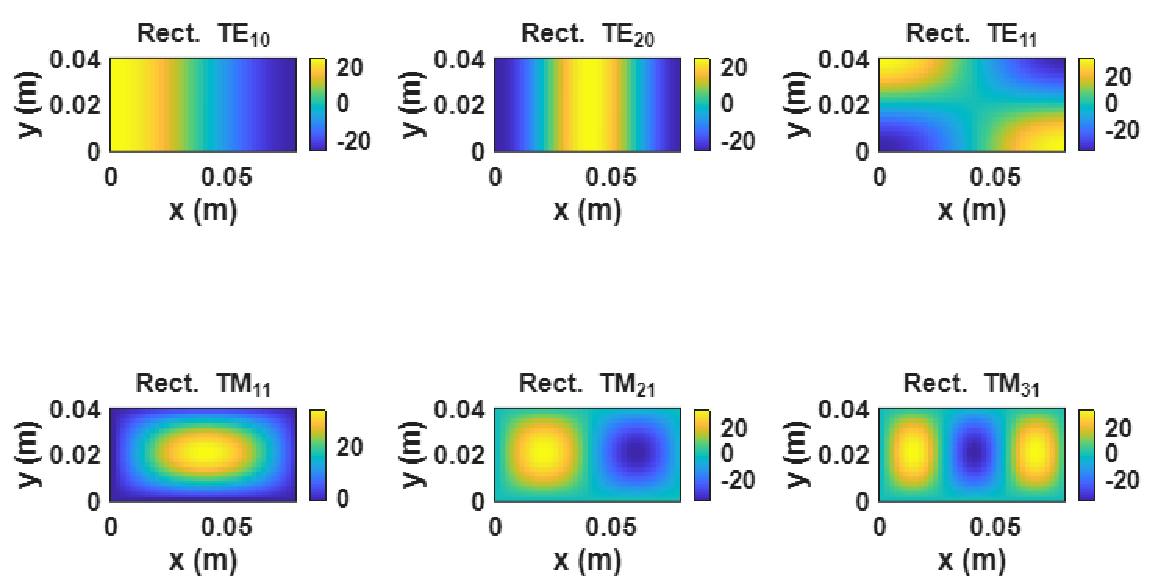}
    \caption{Scalar modal field distributions for the rectangular waveguide.}
    \label{fig:rect_scalar_modes}
\end{figure}

\begin{figure}
    \centering
    \includegraphics[width=\linewidth]{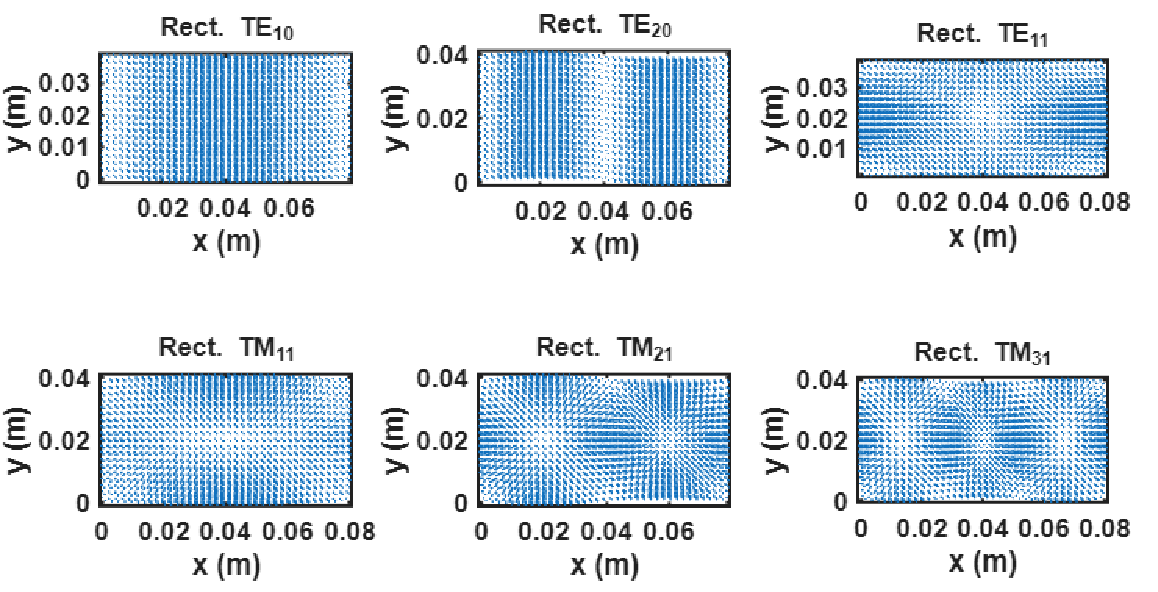}
    \caption{Reconstructed transverse electric field distributions for the rectangular waveguide.}
    \label{fig:rect_transverse_quiver}
\end{figure}

Figure~\ref{fig:rect_scalar_modes} shows the scalar modal field distributions, while Fig.~\ref{fig:rect_transverse_quiver} shows the reconstructed transverse electric field distributions using quiver plots. Together, these plots confirm the expected modal ordering, cutoff behavior, and field symmetry for the rectangular guide.

\subsection{Validation for the Circular Waveguide}

\begin{figure}
    \centering
     \includegraphics[width=0.90\linewidth]{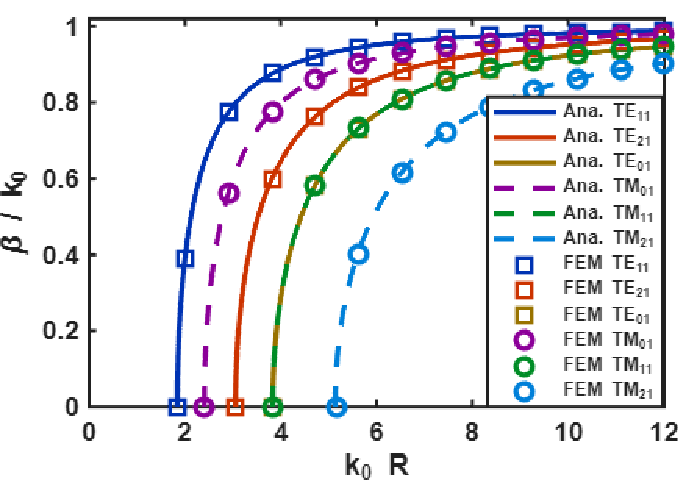}
    \caption{Analytical and FEM dispersion curves for the circular waveguide.}
    \label{fig:circ_dispersion}
\end{figure}

The circular waveguide was used as the second validation case because its modal cutoff values are also available analytically, but now through Bessel-function roots rather than Cartesian separation constants. This case therefore provides a second benchmark while also testing the FEM solver on a curved PEC boundary.

For the circular waveguide, the analytical cutoff relations are written in terms of \(J_n(x)\) denoting the Bessel function of the first kind of order \(n\). For TE modes,
\begin{equation}
k_{c,nm}^{(\mathrm{TE})}
=
\frac{p_{nm}'}{R},
\label{eq:circ_te_kc_num}
\end{equation}
and
\begin{equation}
f_{c,nm}^{(\mathrm{TE})}
=
\frac{p_{nm}'}{2\pi R\sqrt{\mu\epsilon}},
\label{eq:circ_te_fc_num}
\end{equation}
where $p_{nm}'$ is the $m$th root of $J_n'(x)$. For TM modes,
\begin{equation}
k_{c,nm}^{(\mathrm{TM})}
=
\frac{p_{nm}}{R},
\label{eq:circ_tm_kc_num}
\end{equation}
and
\begin{equation}
f_{c,nm}^{(\mathrm{TM})}
=
\frac{p_{nm}}{2\pi R\sqrt{\mu\epsilon}},
\label{eq:circ_tm_fc_num}
\end{equation}
where $p_{nm}$ is the $m$th root of $J_n(x)$.

From these expressions, the dominant circular-waveguide mode is TE$_{11}$, with cutoff frequency
\begin{equation}
f_{c,11}^{(\mathrm{TE})}
=
\frac{p_{11}'}{2\pi R\sqrt{\mu\epsilon}},
\label{eq:circ_te11_fc}
\end{equation}
where $p_{11}'$ is the first root of $J_1'(x)$. There is no TE$_{10}$ mode because the radial index must satisfy $m\ge 1$, although TE$_{01}$ does exist. For TM modes, the first propagating mode is TM$_{01}$, with cutoff frequency
\begin{equation}
f_{c,01}^{(\mathrm{TM})}
=
\frac{p_{01}}{2\pi R\sqrt{\mu\epsilon}},
\label{eq:circ_tm01_fc}
\end{equation}
where $p_{01}$ is the first root of $J_0(x)$. There is also no TM$_{10}$ mode.

Compared with the rectangular guide, the circular guide required an unstructured triangular mesh so that the curved PEC wall could be represented more naturally. As in the rectangular case, the FEM results were post-processed to retain only the first three \emph{unique} TE and TM modes. This was done to avoid repeated reporting of numerically identical or nearly identical cutoff roots. In the circular guide, such behavior is especially relevant because certain modes can appear as degenerate pairs due to the rotational symmetry of the geometry.

Table~\ref{tab:circular_cutoff_comparison} compares the analytical and FEM cutoff values for the first three unique TE modes and the first three unique TM modes of the circular waveguide.

\begin{table}[t]
\centering
\caption{Circular-waveguide cutoff comparison}
\label{tab:circular_cutoff_comparison}
\begin{tabular}{ccccc}
\toprule
Mode & Analytical $k_c$ & FEM $k_c$ & Analytical $f_c$ & FEM $f_c$ \\
     & (rad/m) & (rad/m) & (GHz) & (GHz) \\
\midrule
TE$_{11}$ & 61.372793 & 61.437593 & 2.930335 & 2.933429 \\
TE$_{21}$ & 101.807898 & 102.139219 & 4.860969 & 4.876788 \\
TE$_{01}$ & 127.723532 & 128.643771 & 6.098349 & 6.142288 \\
TM$_{01}$ & 80.160852 & 80.325116 & 3.827399 & 3.835242 \\
TM$_{11}$ & 127.723532 & 128.161641 & 6.098349 & 6.119268 \\
TM$_{21}$ & 171.187410 & 172.928669 & 8.173597 & 8.256736 \\
\bottomrule
\end{tabular}
\end{table}

Figure~\ref{fig:circ_dispersion} shows the analytical and FEM dispersion curves for the circular guide. The FEM results again track the analytical behavior closely.

\begin{figure}
    \centering
    \includegraphics[width=\linewidth]{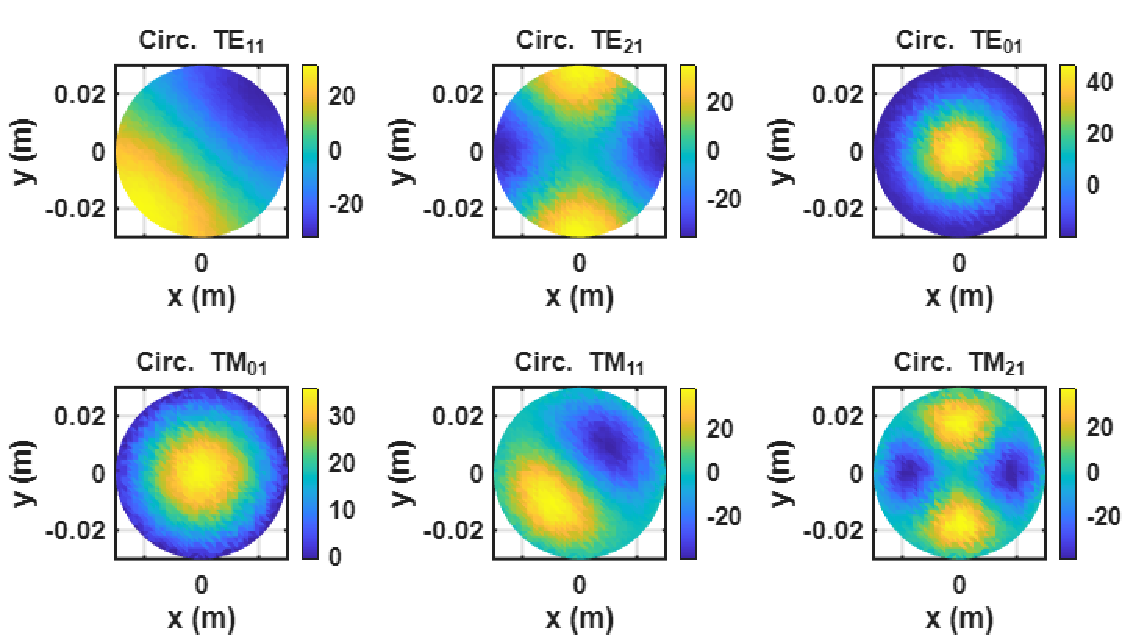}
    \caption{Scalar modal field distributions for the circular waveguide.}
    \label{fig:circ_scalar_modes}
\end{figure}

\begin{figure}
    \centering
    \includegraphics[width=\linewidth]{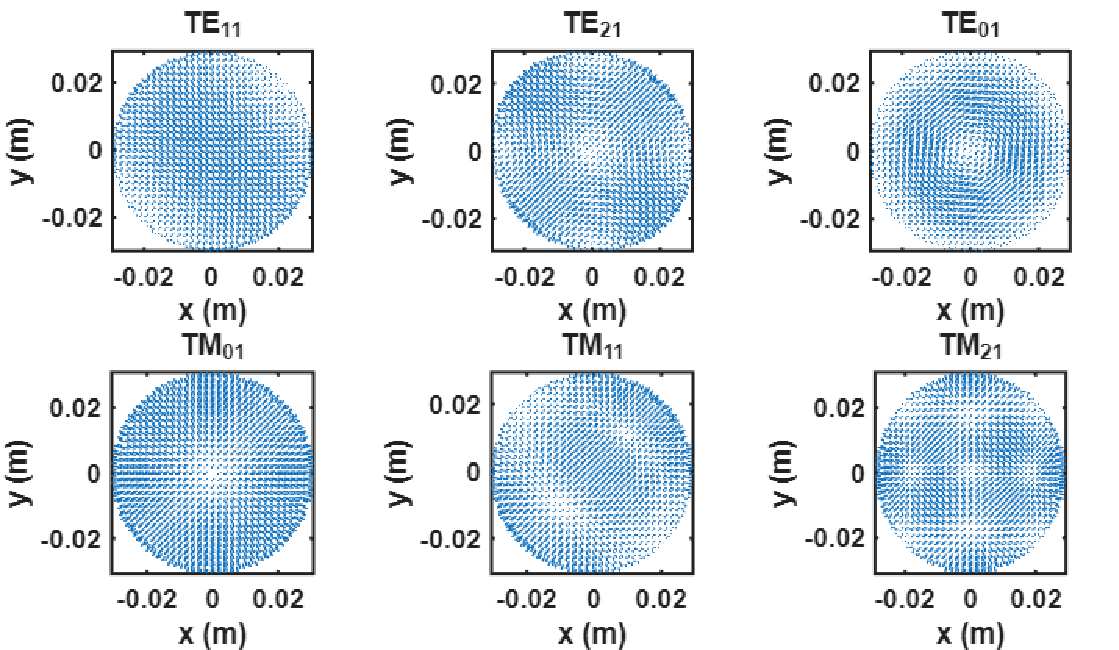}
    \caption{Reconstructed transverse electric field distributions for the circular waveguide.}
    \label{fig:circ_transverse_quiver}
\end{figure}

Figure~\ref{fig:circ_scalar_modes} shows the scalar modal distributions, and Fig.~\ref{fig:circ_transverse_quiver} shows the reconstructed transverse electric field patterns. The dominant TE$_{11}$ mode and the retained higher-order TE and TM modes show the expected circular symmetry and nodal structure.

\subsection{Ridged-Waveguide Results}
\begin{figure}[t]
    \centering
    \includegraphics[width=0.75\linewidth]{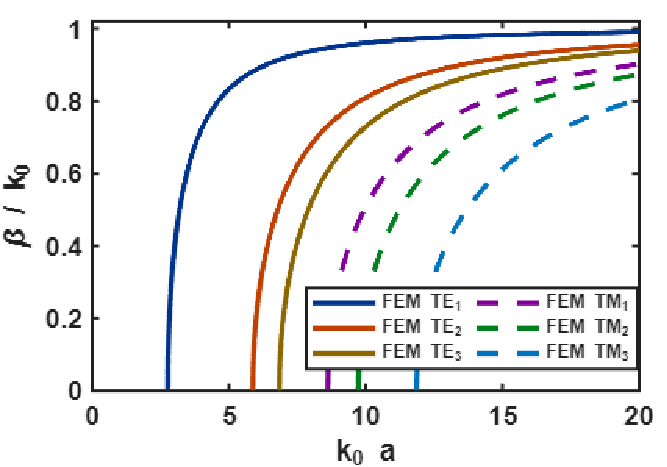}
    \caption{FEM dispersion curves for the single-ridged waveguide.}
    \label{fig:single_ridge_dispersion}
\end{figure}

\begin{figure}
    \centering
    \includegraphics[width=0.75\linewidth]{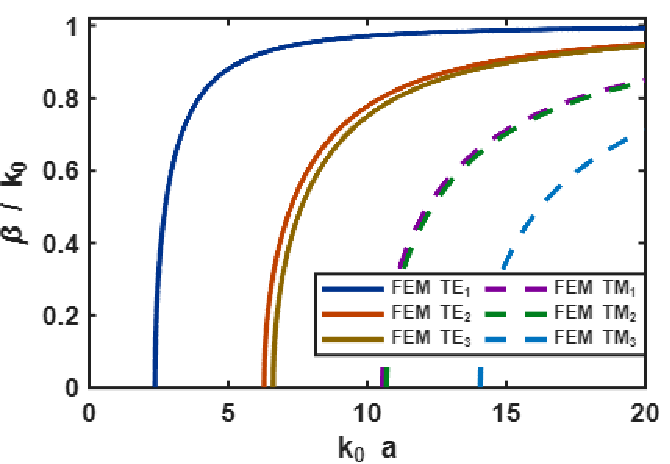}
    \caption{FEM dispersion curves for the double-ridged waveguides.}
    \label{fig:double_ridge_dispersion}
\end{figure}

After validating the solver using the rectangular and circular guides, the same FEM procedure was applied to the single-ridged and double-ridged waveguides. From a numerical point of view, the main extra step in the ridge cases was the geometry construction itself. Unlike the empty rectangular and circular guides, the air-filled computational domain had to be built by excluding the metallic ridge region and treating the ridge surfaces as PEC boundaries during meshing and assembly. The same unique-mode post-processing was also used here so that closely spaced or repeated cutoff roots were not plotted redundantly.

Figure~\ref{fig:single_ridge_dispersion} and Figure~\ref{fig:double_ridge_dispersion} shows the FEM dispersion curves for the single-ridged and double-ridged waveguides.

\begin{figure}
    \centering
    \includegraphics[width=\linewidth]{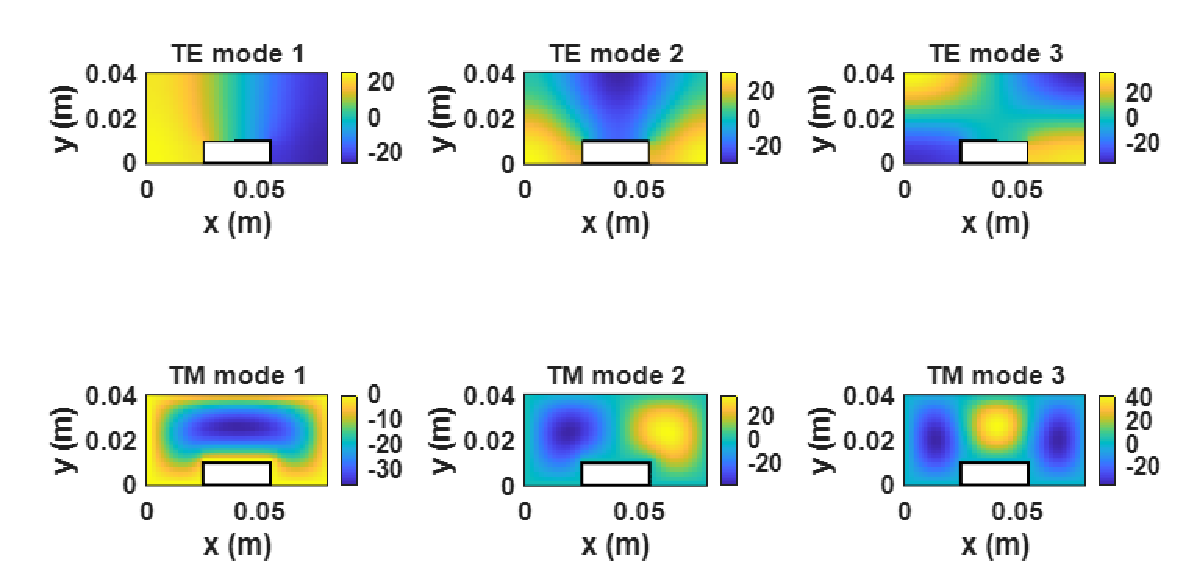}
    \caption{Scalar modal fields  for the single-ridged waveguide.}
    \label{fig:single_ridge_fields}
\end{figure}

\begin{figure}
    \centering
    \includegraphics[width=\linewidth]{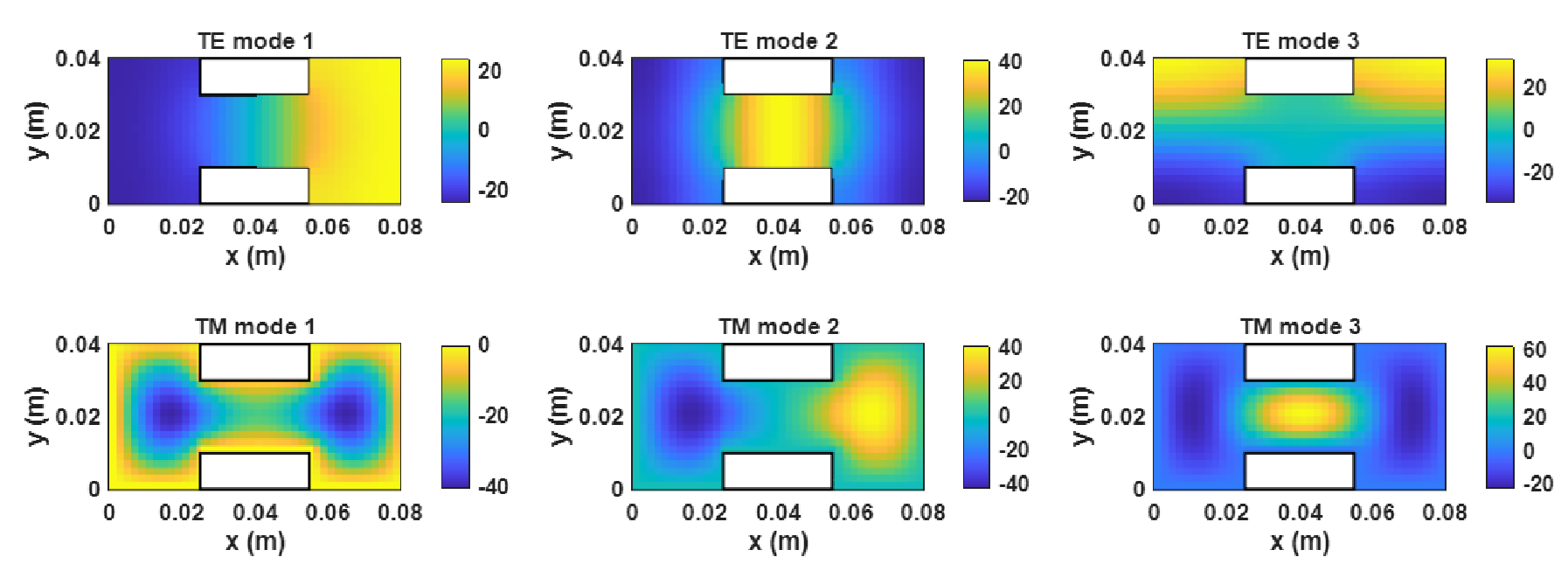}
    \caption{Scalar modal fields for the double-ridged waveguide.}
    \label{fig:double_ridge_fields}
\end{figure}

\begin{figure}
    \centering
    \includegraphics[width=\linewidth]{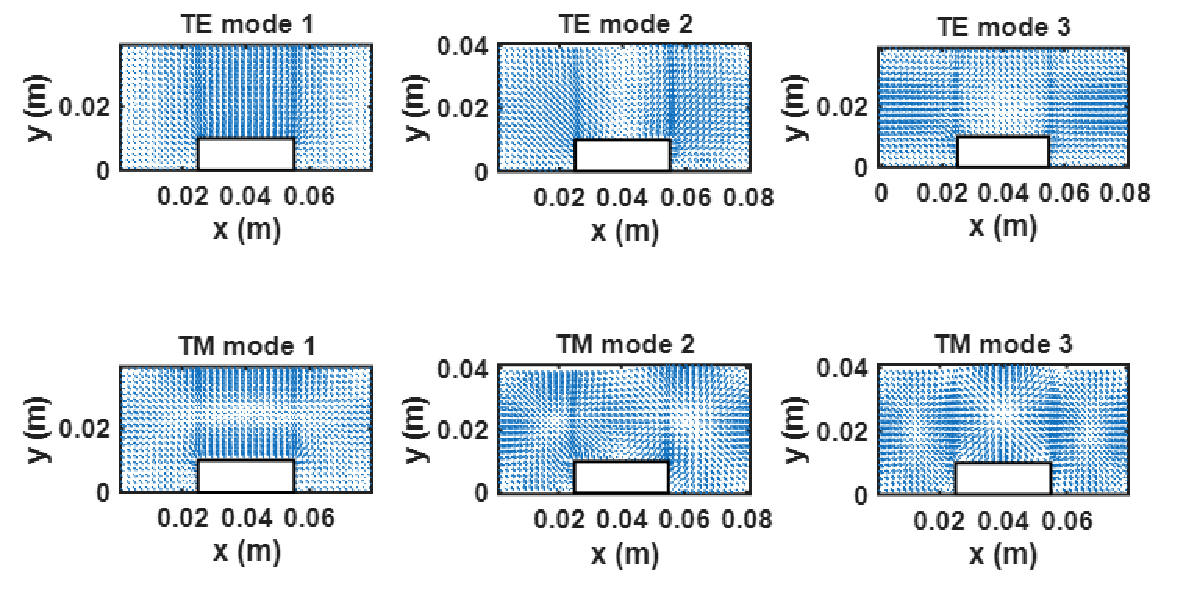}
    \caption{Transverse electric field plots for the single-ridged waveguide.}
    \label{fig:single_ridge_fields_quiver}
\end{figure}

\begin{figure}
    \centering
    \includegraphics[width=\linewidth]{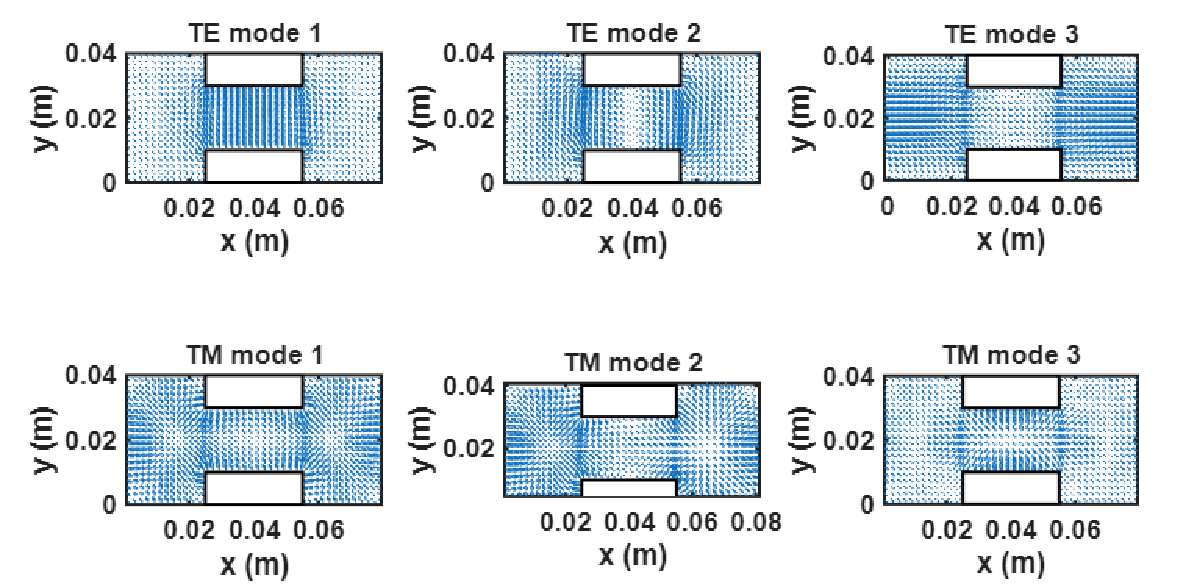}
    \caption{Transverse electric field plots for the double-ridged waveguide.}
    \label{fig:double_ridge_fields_quiver}
\end{figure}

Figures~\ref{fig:single_ridge_fields} and \ref{fig:double_ridge_fields} show the corresponding scalar modal fields. Figures~\ref{fig:single_ridge_fields_quiver} and \ref{fig:double_ridge_fields_quiver} show reconstructed transverse electric field distributions. For the double-ridged waveguide in Figure~\ref{fig:double_ridge_dispersion} we can see that the TM$_1$ and TM$_2$ branches remain very close to each other even after the unique-mode filtering. This indicates a near-degenerate physical behavior of the geometry rather than a simple repeated numerical root, so the two branches were retained in the results.

\begin{figure}[t]
    \centering
    \includegraphics[width=\linewidth]{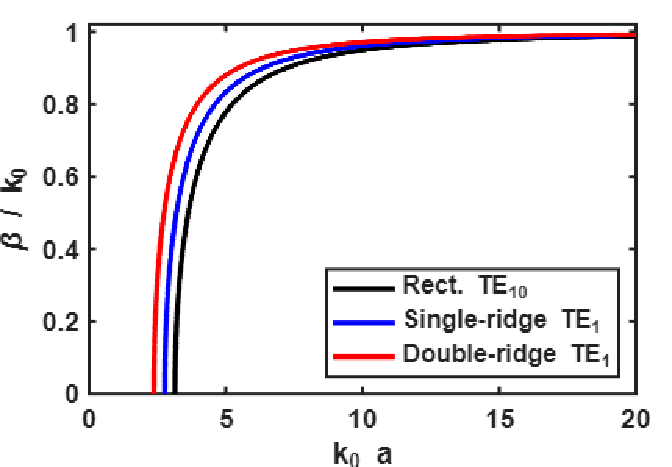}
    \caption{Comparison of dispersion curves for the empty rectangular, single-ridged, and double-ridged waveguides.}
    \label{fig:rect_vs_ridge_dispersion}
\end{figure}

\subsection{Comparison of Rectangular and Ridged Dispersion Curves}

To make the effect of ridge loading more visible, Fig.~\ref{fig:rect_vs_ridge_dispersion} compares the dispersion curves of the empty rectangular, single-ridged, and double-ridged waveguides on the same axes. The main trend is the reduction of the dominant-mode cutoff when ridges are introduced. The single-ridged guide lowers the cutoff relative to the empty rectangular guide, while the double-ridged guide produces an even larger shift. This comparison directly illustrates the effect of modifying the rectangular cross section by ridge loading.

\section{Conclusion}

A 2D finite element method solver was developed to compute modal field distributions and dispersion characteristics of empty metallic waveguides. The formulation was based on a scalar Helmholtz eigenvalue problem over the transverse cross section, with Dirichlet boundary conditions used for TM modes and Neumann boundary conditions used for TE modes. Using linear triangular elements, the method produced a generalized eigenvalue problem whose eigenvalues correspond to the cutoff wavenumbers of the supported modes.

The implementation was first validated on empty rectangular and circular waveguides. These two cases provide reliable benchmarks because their cutoff values and modal structures are known analytically. The FEM results were then extended to single-ridged and double-ridged waveguides, where the geometry is more complicated and the benefit of the numerical method becomes more apparent. In the ridged cases, the dominant-mode cutoff decreased and the modal fields became more concentrated near the ridge region, which is consistent with physical expectations.

Overall, the results show that the FEM approach used here provides a flexible and reliable tool for waveguide modal analysis. Several improvements are possible in future work. Higher-order basis functions could be used to improve accuracy, especially near curved boundaries and ridge corners. Additional mesh refinement studies could be performed to quantify convergence more carefully. The same framework could also be extended to partially filled or dielectric-loaded guides, where analytical solutions are less accessible and numerical eigenmode analysis becomes even more valuable.

\end{document}